\documentclass[prd,twocolumn,superscriptaddress,amsfonts,amssymb,amsmath,showpacs]{revtex4-2}

\usepackage[utf8]{inputenc}
\usepackage[T1]{fontenc}  
\usepackage{lmodern}      

\usepackage{amsmath, amssymb, amsfonts, latexsym, bm}
\usepackage{graphicx, epsfig, float, dcolumn, booktabs}
\usepackage{ragged2e, textcomp, xcolor}
\usepackage{commath}
\usepackage{orcidlink}

\usepackage[sc]{mathpazo}
\usepackage{subcaption}
\usepackage{caption}

\usepackage{hyperref}
\hypersetup{
  colorlinks=true,
  linkcolor=blue,
  citecolor=blue,
  urlcolor=blue
}

\usepackage{cleveref}

\allowdisplaybreaks[1]

\begin{document}

\color{black}       

\title{Late time behavior in $f(R,\mathcal{L}_{m})$ gravity through Gaussian reconstruction and dynamical stability}
\author{Y. Kalpana Devi\orcidlink{0009-0001-2686-7281}}
\email{kalpanayengkhom123@gmail.com}
\affiliation{Department of Mathematics,
Birla Institute of Technology and Science-Pilani, Hyderabad Campus, Jawahar Nagar, Kapra Mandal, Medchal District, Telangana 500078, India.}

\author{S.A. Narawade\orcidlink{0000-0002-8739-7412}}
\email{shubhamn2616@gmail.com}

\affiliation{Department of Mathematics,
Birla Institute of Technology and Science-Pilani, Hyderabad Campus, Jawahar Nagar, Kapra Mandal, Medchal District, Telangana 500078, India.}

\author{B. Mishra\orcidlink{0000-0001-5527-3565}}
\email{bivu@hyderabad.bits-pilani.ac.in}
\affiliation{Department of Mathematics,
Birla Institute of Technology and Science-Pilani, Hyderabad Campus, Jawahar Nagar, Kapra Mandal, Medchal District, Telangana 500078, India.}


\begin{abstract}
\textbf{Abstract}: In this paper, we explore modified gravity in the framework of $f(R, \mathcal{L}_m)$ theories by reconstructing the function $f(\mathcal{L}_m)$, where $\mathcal{L}_m = \rho$ is the matter Lagrangian, under the assumption of a pressureless, matter-dominated Universe. Using a non-parametric Gaussian process reconstruction technique applied to Hubble data, we obtain two viable models of $f(\mathcal{L}_m)$ : (i) a power-law model $f_1(\mathcal{L}_m) = \alpha \mathcal{L}_m^{b_1}$ with $b_1 \in [0.018, 0.025]$ and (ii) an exponential model $f_2(\mathcal{L}_m) = \alpha \mathcal{L}_{m0} \left(1 - e^{-b_2 \sqrt{\mathcal{L}_m/\mathcal{L}_{m0}}} \right)$ with $b_2 \in [2.3, 3.0]$. We then fix the parameter values within these reconstructed ranges and analyze the corresponding dynamical systems within the matter-dominated epoch by constructing autonomous equations. Phase-space analysis reveals the presence of stable critical points in both models, suggesting viable cosmic evolution within their domains of validity. Both the models exhibit stable attractor solution at late time, reinforcing their viability in explaining the late time cosmic acceleration without explicitly invoking a cosmological constant. Our results indicate that $f(R, \mathcal{L}_m)$ gravity with data-driven matter-sector modifications can offer a compelling alternative description of cosmic dynamics during the matter-dominated era.
\end{abstract}

\maketitle

\section{Introduction} 
One of the most significant milestones in modern cosmology has been the discovery of the accelerated expansion of the Universe. This phenomenon was initially observed through Type Ia Supernovae \cite{Riess_1998, Perlmutter_1999} and later corroborated by precise measurements of the Cosmic Microwave Background (CMB) \cite{Spergel_2007, Komatsu_2011} and Baryon Acoustic Oscillations (BAO) \cite{Eisenstein_2005,Percival_2010}. The growing body of observational evidence has solidified the accelerated expansion as a well-established feature of our Universe and has prompted a deeper inquiry into its fundamental causes \cite{Weinberg_2013}. To account for this accelerated behavior, various theoretical approaches have been proposed. Einstein's General Relativity (GR) has been the heart of modern cosmology, it has been successful in describing gravitational phenomena on solar system scales and the large-scale structure of the Universe. But GR face challenges in explaining the late-time cosmic acceleration of the Universe. To account for this acceleration, the standard 
$\Lambda$CDM model introduces a cosmological constant 
$\Lambda$, interpreted as dark energy \cite{Weinberg1989}. Despite its empirical success \cite{Wilson2006,Davis2007}, $\Lambda$CDM faces theoretical issues, such as the fine-tuning \cite{Weinberg1989}, coincidence problem \cite{Steinhardt1999} and age problem \cite{Copeland2006,Padmanabhan2003}, prompting the exploration of alternative explanations. 

Modified gravity theories has been proven highly effective in explaining the late-time acceleration of the Universe \cite{CAPOZZIELLO2011167,NOJIRI201159}. Among the most widely studied approaches is $f(R)$ gravity, where the standard Einstein-Hilbert action of GR is generalized by replacing the Ricci scalar $R$ with a function $f(R)$. This modification introduces new dynamical degrees of freedom and can naturally account for the observed cosmic acceleration without resorting to a cosmological constant \cite{Sotiriou2010}. The theoretical flexibility of 
$f(R)$ models makes them attractive candidates for addressing the fine-tuning and coincidence problems inherent in the standard $\Lambda$CDM paradigm. A further generalization of $f(R)$ gravity is the $f(R,\mathcal{L}_{m})$ gravity, where the action depends on both the Ricci scalar $R$ and the matter Lagrangian $\mathcal{L}_{m}$  \cite{Harko2010}. These theories introduce a non-minimal coupling between geometry and matter, leading to the non-conservation of the energy-momentum tensor and the emergence of an extra force acting on massive particles. As a result, particle motion deviates from geodesic paths, offering rich phenomenology and potential signatures of new physics. 

A fundamental challenge in modified theories of gravity has been the arbitrary selection of the functional form of the gravity model. To develop a modified gravity model that is both observationally consistent and theoretically viable, the model must be consistent with various cosmological observations and also free from theoretical inconsistencies like instabilities, ghost degrees of freedom or violations of fundamental conservation laws. Early attempts to reconstruct cosmic acceleration often relied on simple parametric form of the function, such as the Chevallier-Polarski-Linder (CPL) parametrization for the dark-energy equation of state \cite{Chevallier2001,Linden2008} or on cosmographic series expansions of the scale factor in powers of redshift \cite{Visser2004}. While these methods are straightforward and computationally light, they suffer from  major drawbacks such as the choice of parametrization can bias the outcome and series expansions often break down at moderate to high redshift. In contrast, Gaussian Process (GP) regression provides a fully probabilistic, non-parametric framework that treats the function itself as a distribution conditioned only on the data \cite{Rasmussen2005,Seikel2013,Rasmussen2005}. GP allow us to obtain the form of the involved function in a model independent way only from the datasets. This method has been used to study the expansion history of the Universe and various dark energy models using different data sets \cite{Seikel2013,Seikel_2012,Shafieloo2014,Kim2013,Nair2014,Yang2015,GomezValent2018,Cai2020}.

In this study, we will perform GP analysis for the case of $f(R,\mathcal{L}_{m})$ gravity within the framework of FLRW metric in order to reconstruct the cosmologically viable functional form of $f(R,\mathcal{L}_{m})$ in a model-independent way by using observational values of Hubble measurements $H(z)$. In Sec. \ref{sec2}, we have provided a brief description of the $f(R,\mathcal{L}_{m})$ gravity from action integral till the Friedmann equations. It is followed by the reconstruction of Hubble parameter $H(z)$ and its derivative $H'(z)$ through GP using various observational dataset in Sec. \ref{sec3}. In Sec. \ref{sec4}, we reconstructed the functional form of $f(\mathcal{L}_{m})$ and obtained two viable cosmological models that lie within the $1\sigma$ uncertainty region. We performed the stability analysis of the obtained models using dynamical system analysis in Sec. \ref{sec5} and the results are presented and discussed in Sec. \ref{sec6}.

\section{\texorpdfstring{$f(R,\mathcal{L}_m)$}{} gravity framework}\label{sec2}

The action of $f(R,\mathcal{L}_m)$ gravity can be defined as,
\begin{equation}\label{cfrl1}
    S=\int d^4x \sqrt{-g}f(R,\mathcal{L}_m)~,
\end{equation}
in which $g$ is the metric determinant and in units such that $8\pi G$ also equals $1$, which will be assumed and $f(R,\mathcal{L}_{m})$ is an arbitrary function of Ricci scalar $R$ and matter Lagrangian $\mathcal{L}_m$. The value of Ricci scalar can be obtained by contracting Ricci tensor as,
\begin{align}\label{Ricci_scalar}
    R=g^{\mu \nu} R_{\mu \nu}~.
\end{align}
For a positive metric signature, the value of Ricci tensor is defined by using the equation,
\begin{equation}
    R_{\mu \nu}=\frac{\partial \Gamma^{\beta}_{\mu \nu}}{\partial x^{\beta}}- \frac{\partial \Gamma^{\beta} _{\mu \beta}}{\partial x^\nu}+\Gamma^{\alpha}_{\mu \nu} \Gamma^{\beta}_{\alpha \beta}-\Gamma^{\alpha}_{\mu \beta} \Gamma^{\beta}_{\nu \alpha}~,
\end{equation}    
By applying the variational principle in the above action, one may obtain the $f(R,\mathcal{L}_m)$ gravity field equations as,
\begin{align}\label{Eq:field1}
      R_{\mu\nu}f_R+(g_{\mu\nu}\nabla_\mu\nabla^\mu-\nabla_\mu\nabla_\nu )f_R&-\frac{f}{2}g_{\mu\nu}\nonumber\\
      &=\frac{1}{2}f_{\mathcal{L}_m}(T_{\mu\nu}-\mathcal{L}_mg_{\mu\nu})~,  
\end{align}
for which $R_{\mu\nu}$ is the Ricci tensor, $g_{\mu\nu}$ is the metric tensor, $f_R=\frac{\partial f(R,~\mathcal{L}_m)}{\partial R}$ and $f_{\mathcal{L}_m}=\frac{\partial f(R,~\mathcal{L}_m)}{\partial \mathcal{L}_m}$. For $f(R,\mathcal{L}_{m})=f_{1}(R)+f_{2}(R)G(\mathcal{L}_{m})$, where $f_{1}(R)$ and $f_{2}(R)$ are arbitrary Ricci scalar functions and $G(\mathcal{L}_{m})$ is a function of matter Lagrangian density. For $f_{1}(R) = \frac{R}{2}$, $f_{2}(R)=1$ and $G(\mathcal{L}_{m})=\mathcal{L}_{m}$, we recover the standard field equation for GR and when $f_{2}(R)=1$ and $G(\mathcal{L}_{m})=\mathcal{L}_{m}$ it reduces to $f(R)$ gravity. Moreover, for $G(\mathcal{L}_{m})=1+\lambda \mathcal{L}_{m}$ matter and geometry are linearly coupled.

The contracting form of the field equation \eqref{Eq:field1} can be written as
\begin{equation}\label{Eq:field2}
f_{R}R+3\Box f_{R}-2f=f_{\mathcal{L}_{m}}\left(\frac{1}{2}T-2\mathcal{L}_{m}\right)~.  
\end{equation}
By eliminating the term $\Box f_{R}$ between Eqs. \eqref{Eq:field1} and \eqref{Eq:field2}, we obtain
\begin{align}
  f_{R}\left(R_{\mu \nu}-\frac{1}{3}R g_{\mu \nu}\right)&+\frac{f}{6}-\nabla_{\mu}\nabla_{\nu}f_{R}\nonumber\\
 & =\frac{1}{2}f_{\mathcal{L}_{m}}\left(T_{\mu \nu}-\frac{1}{3}(T-\mathcal{L}_{m}) g_{\mu \nu} \right)~.  
\end{align}
Taking the covariant divergence of Eq. \eqref{Eq:field1}, with the use of the following mathematical identity
\begin{equation}
\nabla^{\mu}\left[R_{\mu \nu }f_{R}+( g_{\mu\nu}\Box-\nabla_{\mu}\nabla_{\nu})f_{R}-\frac{f}{2}g_{\mu \nu}\right]=0~, 
\end{equation}
we can obtain the divergence of the energy-momentum tensor $T_{\mu \nu}$ as,
\begin{eqnarray}
\nabla^{\mu}T_{\mu \nu}&=&\nabla^{\mu}\ln[f_{\mathcal{L}_{m}}](\mathcal{L}_{m}g_{\mu \nu}-T_{\mu \nu}) \nonumber \\
&=&2\nabla^{\mu}\ln[f_{\mathcal{L}_{m}}]\frac{\partial \mathcal{L}_{m}}{\partial g^{\mu \nu}}~.
\end{eqnarray}
The requirement of the conservation of the energy-momentum tensor of matter,
i.e., $\nabla^{\mu}T_{\mu \nu}=0$, gives an effective functional relation between the matter and Lagrangian density. Moreover, the conservation of the energy-momentum tensor yields
\begin{equation}\label{cfrl3}
	\nabla^\mu\ln f_{\mathcal{L}_{m}}=0~.
\end{equation}
We assume the Universe to be defined by the flat FLRW space-time and the stress energy-momentum tensor for a perfect fluid which can be written as,
\begin{align}\label{stressenergy}
    T_{\mu \nu}=(p + \rho ) u_{ \mu } u_{ \nu }+ p g_{ \mu \nu }~,
\end{align}
where $u_{\mu}$ is the four velocity vectors along the time directions, $\rho $ is the matter energy density and $p$ is the isotropic pressure. On solving the field equation in Eq. \eqref{Eq:field1} we get the Friedmann equations for $f(R, \mathcal{L}_m)$ gravity that describes the dynamics of the Universe as,
\begin{eqnarray}\label{eqn1}
    3H^2f_R+\frac{1}{2}(f-f_R R-f_{\mathcal{L}_m}\mathcal{L}_{m})+3H\dot{f}_{R}=\frac{1}{2}f_{\mathcal{L}_m}\rho~,\nonumber\\ 
    \dot{H}f_R+3H^2f_R-\ddot{f_R}-3H\dot{f_R}+\frac{1}{2}(f_{\mathcal{L}_m}\mathcal{L}_{m}-f)=\frac{1}{2}f_{\mathcal{L}_m}p~,\nonumber\\
        \label{eqn2}
\end{eqnarray}
where $H=\frac{\dot{a}}{a}$ is the Hubble parameter. It gives us the present day expansion rate of the Universe and also helps us in understanding the age, size and evolution of the Universe. The above equations gives us a relation between density parameter for various components of the Universe and Hubble parameter for a homogeneous and isotropic Universe. Understanding the hubble parameter as a function of redshift $z$ can help us understand the cosmic expansion of the Universe. 

\section{Gaussian Process}\label{sec3}
GP is a powerful non-parametric statistical technique used for regression and data reconstruction. It provides a probabilistic approach to reconstruct the behavior of a function and its derivatives from observational data without assuming a specific parametric form. In this work we have used publicly available code GaPP (Gaussian Processes in Python) which is described in \cite{Seikel_2012}.
\subsection{GP}
GP is a generalization of gaussian distribution, given a set of $N$ gaussian distributed discrete data points we can construct a continuous function along with its confidence level which describes the data set without assuming any functional form. For $N$ observational data points we have  $y = (y_1, \dots, y_N)$ at redshifts $Z = (z_1, \dots, z_N)$ with the errors in observation is given by the covariance matrix $C$. We can reconstruct the function at the data points $Z^* = (z^*_1, \dots, z^*_N)$ and denote the function values at these points as $f^* = (f(z^*_1), \dots, f(z^*_N))$ which are described by normal distribution. The function values at point $z$ and $z^*$ are dependent on each other and the relation between them is given by the covariance function $C(z^*, \tilde{z}^*)$. As GP is a generalization of gaussian distribution so we can express it as a distribution over function which is described by mean function $\mu(z)$ and covariance matrix $C(z,\tilde{z})$ which is written as :
\begin{equation}\label{gp}
    f(z) \sim \mathcal{GP} (\mu(z), C(z, \tilde{z})).
\end{equation}
From Eq. \eqref{gp} we can generate a continuous GP curve, in order to reconstruct at other redshifts for which we do not have the data i.e at $Z^{*}$ we need to define a specific covariance function which is also called as the kernel function in GP, $C(z^{*},\tilde{z^{*}})=K(z^{*},\tilde{z^{*}})$. The kernel plays an important role in the GP reconstruction, it give the strength of correlation between reconstructed points and their deviation from the mean values. In this work we will be using the general Square Exponential kernel 
\begin{equation}
     K(z, \tilde{z}) = \sigma_f^2 \exp\left(-\frac{(z - \tilde{z})^2}{2\ell^2}\right)~.
\end{equation}
Here $'{\sigma_f}'$ and $'\ell '$ are the hyperparameters which describes the correlation between the reconstructed data points, the length scale $'\ell '$ characterizes the distance it has to travel to obtain significant change in $x$-direction and $'{\sigma_{f}}'$ denotes the change in $y$-direction. The Square Exponential kernel function is infinitely differentiable, more smooth function and decays exponentially.

\subsection{Reconstruction of $H(z)$ from observational data} 
In this work, we apply GP together with the squared exponential in $H(z)$ data from cosmic chronometers (CC), supernovae of Type Ia (SN) and baryonic acoustic oscillation (BAO). For CC data set, we have considered data points at redshift $z \lesssim 2$ from \cite{Moresco_2016}. For the SN data, we used compressed pantheon compilation data extracted from the supernovae of Type Ia combined with the Hubble Space Telescope (HST) CANDLES and CLASH  Multi-cycle Treasury data (MCT) from \cite{Riess_2018} in which we get five distinct data points as for the data point at redshift $z=1.5$ is not gaussian distributed as shown in \cite{GomezValent_2018}. First we apply GP to CC data only and construct $H(z)$ at $z=0$ and through this obtained $H_{0}$ value in SN data along this the correlation matrix we can obtain $H(z)=E(z)H_{0}$, then we apply the GP reconstruction to the CC+SN data succesively until the obtained $H_{0}$ value and its uncertainty converges to $\lesssim 10^{-4}$. For  BAO data, we have considered the ten $H(z)$ data points from the radial BAO method as in \cite{Zhang_2016}. The data points are shown in the table below \ref{tab:HubbleData} and \ref{tab:EzCorrMatrix}.

\begin{table}[ht]
\centering
\scriptsize
\resizebox{\columnwidth}{!}{%
\begin{tabular}{|ccc|ccc|}
\hline
\textbf{z} & \textbf{H(z) [km/s/Mpc]} & \textbf{$\sigma_H$ [km/s/Mpc]} & \textbf{z} & \textbf{H(z) [km/s/Mpc]} & \textbf{$\sigma_H$ [km/s/Mpc]} \\
\hline
0.07   & 69.0  & 19.6  & 0.4783 & 80.9  & 9.0\\
0.09   & 69.0  & 12.0  & 0.48   & 97.0  & 62.0 \\
0.12   & 68.6  & 26.2  & 0.593  & 104.0 & 13.0 \\
0.17   & 83.0  & 8.0   & 0.680  & 92.0  & 8.0  \\
0.179  & 75.0  & 4.0   & 0.781  & 105.0 & 12.0\\
0.199  & 75.0  & 5.0   & 0.875  & 125.0 & 17.0 \\
0.200  & 72.9  & 29.6  & 0.880  & 90.0  & 40.0 \\
0.27   & 77.0  & 14.0  & 0.900  & 117.0 & 23.0 \\
0.280  & 88.8  & 36.6  & 1.037  & 154.0 & 20.0 \\
0.352  & 83.0  & 14.0  & 1.300  & 168.0 & 17.0 \\
0.3802 & 83.0  & 13.5  & 1.363  & 160.0 & 33.6 \\
0.4    & 95.0  & 17.0  & 1.430  & 177.0 & 18.0 \\
0.4004 & 77.0  & 10.2  & 1.530  & 140.0 & 14.0 \\
0.4247 & 87.1  & 11.2  & 1.750  & 202.0 & 40.0 \\
0.44497 & 92.8 & 12.9  & 1.965  & 186.5 & 50.4 \\
\hline
\rule{0pt}{0.7cm}
0.24 & 79.69 & 2.65 & 0.60 & 87.9 & 6.1\\
0.35 & 84.4 & 7 & 0.73 & 97.3 & 7.0 \\
0.43 & 86.45 & 3.68 & 2.30 & 224 & 8 \\
0.44 & 82.6 & 7.8 & 2.34 & 222 & 7 \\
0.57 & 92.4 & 4.5 & 2.36 & 226 & 8 \\
\hline
\end{tabular}
}
\caption{Hubble parameter measurements from Cosmic Chronometer for the first 30 points as shown in \cite{Moresco_2016} and the bottom 10 points from radial BAO method as shown in \cite{Zhang_2016}  .}
\label{tab:HubbleData}
\end{table}
 
\begin{figure}[ht]
    \centering
    \begin{subfigure}[b]{0.50\textwidth}
        \centering
        \includegraphics[width=88mm]{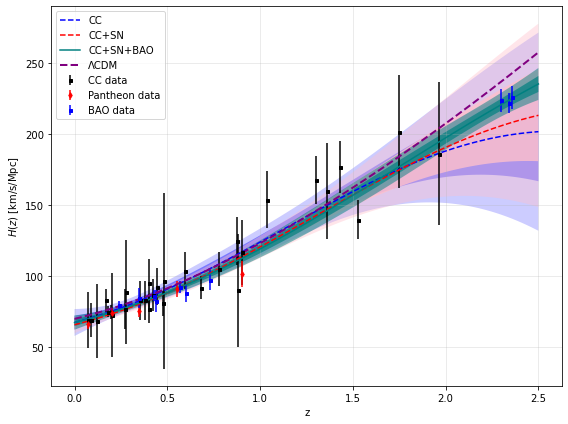}
        \caption{GP reconstruction of $H(z)$.}
        \label{fig:Hz}
    \end{subfigure}%
    \hfill\hfil
    \begin{subfigure}[b]{0.50\textwidth}
        \centering
        \includegraphics[width=88mm]{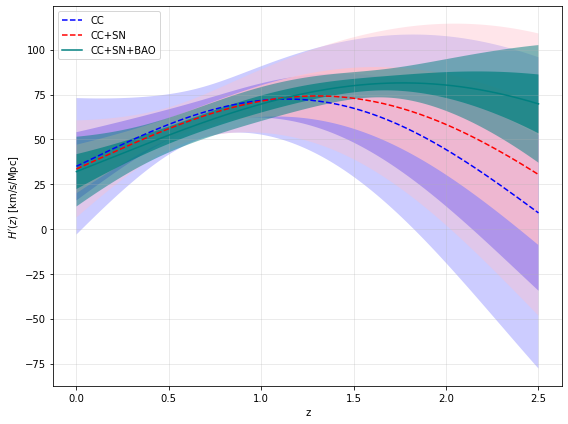}
        \caption{GP reconstruction of $H'(z)$.}
        \label{fig:Hprime}
    \end{subfigure}%
    \caption{Reconstruction  of the Hubble parameter and its first derivative using GP with squared exponential kernel.}
    \label{fig:HHprime}
\end{figure}

\begin{table}[ht]
\centering
\resizebox{\columnwidth}{!}{%
\begin{tabular}{|c|c|ccccc|}
\hline
\textbf{z} & \textbf{$E(z)$} & \multicolumn{5}{c|}{\textbf{Correlation matrix}} \\
\hline
0.07 & $0.997 \pm 0.023$ & 1.00 &       &       &       &       \\
0.20 & $1.111 \pm 0.020$ & 0.39 & 1.00  &       &       &       \\
0.35 & $1.128 \pm 0.037$ & 0.53 & -0.14 & 1.00  &       &       \\
0.55 & $1.364 \pm 0.063$ & 0.37 & 0.37  & -0.16 & 1.00  &       \\
0.90 & $1.520 \pm 0.12$   & 0.01 & -0.08 & 0.17  & -0.39 & 1.00  \\
\hline
\end{tabular}
}
\caption{Values of $E(z)$ and the corresponding correlation matrix taken from \cite{Riess_2018}.}
\label{tab:EzCorrMatrix}
\end{table}
We have reconstructed Hubble as a function of redshift $H(z)$ and its derivative $H'(z)$ using the Hubble data points from various measurements as shown in the above table through GaPP code. The errorbar plot of the GP reconstructed $H(z)$  along with the $1\sigma$ and $2\sigma$ regions is shown in Fig. \ref{fig:Hz} and the $1\sigma$ and $2\sigma$ regions plot of $H'(z)$ is shown in Fig. \ref{fig:Hprime} along with different combination of the data sets: CC, CC+SN and CC+SN+BAO.

\section{Reconstructing the $f(R,\mathcal{L}_{m})$ function from GP}\label{sec4}
In this section, we reconstruct the functional form of $f(R,\mathcal{L}_{m})$  in a cosmological model independent way by using the already reconstructed Hubble parameter and its derivative through GP in the previous section. For this process, we consider the background geometry to be described by the FLRW metric and dominated by $f(R,\mathcal{L}_{m})$ gravity assuming $f(R, \mathcal{L}_{m})=g(R)+f(\mathcal{L}_{m})$, where $g(R)=\frac{R}{2}$ which is the condition where $g(R)$ reduces to GR in $f(R, \mathcal{L}_{m})$ gravity and reconstruct the $f(\mathcal{L}_{m})$. 
We can express the Eq. \eqref{Eq:field1} as a function of redshift $z$ by writing 
\begin{equation}\label{F_lm}
    f_{\mathcal{L}_{m}}=\frac{df(\mathcal{L}_{m})/dz}{d\mathcal{L}_{m}/dz}=\frac{f'(z)}{\mathcal{L}_{m}'(z)}~,
\end{equation}
here, $f'(z)=\frac{df(\mathcal{L}_{m})}{dz}$ and $\mathcal{L}_{m}'(z)$ is the derivative of matter lagrangian $\mathcal{L}_{m}$ with respect to redshift with $(\mathcal{L}_{m}(z)=\rho(z))$ \cite{Harko_2015}, $\mathcal{L}_{m}(z)=\rho(z)=3H_0^2\Omega_{m0}(1+z)^3$ for a matter dominated Universe and $\mathcal{L}_{m}'(z)=6H_0^2\Omega_{m0}(1+z)^2$. In the next step, we will define $f'(z)$ through central differencing method as it has smaller uncertainties than the forward and backward differencing method where
\begin{equation}\label{central differencing}
    f'(z_i) \approx \frac{f(z_{i+1}) - f(z_{i-1})}{z_{i+1} - z_{i-1}}~.
\end{equation}
Using Eq. \eqref{F_lm} and Eq. \eqref{central differencing} in first Friedmann equation as shown in Eq. \eqref{Eq:field1}, we obtain a numerical iterative formula for $f(z)$, given as 
\begin{equation}\label{f(z)iterative}
f(z_{i+1})=f(z_{i-1})+(z_{i+1}-z_{i-1})\frac{2H'(z_i)}{H(z_i)} \left[ \frac{3H(z_i)^2}{2}+\frac{f(z_i)}{2} \right]~.
\end{equation}
We can reconstruct $f(z)$ from Eq. \eqref{f(z)iterative} using the GP reconstructed hubble parameter and its derivative value obtained in section \ref{sec3} using two initial conditions. We get the first initial condition by evaluating the first Friedmann Eq. \eqref{Eq:field1} at present time i.e. $(z=0)$ by imposing $\Lambda$CDM dominated Universe such that $f_{\mathcal{L}_{m}}=0$ and we obtain
\begin{equation}\label{fistcond}
    f(z=0)=6H_0^2(\Omega_{m0}-1)~.
\end{equation}
Here the $H_0$ value is taken from the GP reconstruction and $\Omega_{m0}$ is the value obtained from P18 \cite{Planck_2020} and the second initial condition is evaluated by using forward differencing method through
\begin{equation}
    f'(z_i) \approx \frac{f(z_{i+1}) - f(z_{i})}{z_{i+1} - z_{i}}~,
\end{equation}
and the equation becomes 
\begin{equation}\label{secondcond}
    f(z_{i+1})=f(z_i)+(z_{i+1}-z_{i})\frac{2H'(z_i)}{H(z_i)}\left[\frac{3H(z_i)^2}{2}+\frac{f(z_i)}{2}\right]~.
\end{equation}
Using Eq. \eqref{fistcond} and Eq. \eqref{secondcond} in Eq. \eqref{f(z)iterative}, we can reconstruct the lagrangian $f(z)$ as a function of redshift in a model independent way. Similarly, the matter Lagrangian density  can also be evaluated at each associated redshift using $\mathcal{L}_{m}(z)=3H_0^2\Omega_{m0}(1+z)^3$. So the Lagrangian function $f(\mathcal{L}_{m})$ can be reconstructed in a model independent manner and the errors are propagated through Monte Carlo error propagation to obtain its $1\sigma$ and $2\sigma$ uncertainty region.
\begin{figure}[H]
    \centering
    \begin{subfigure}[b]{0.50\textwidth}
        \centering
        \includegraphics[width=88mm]{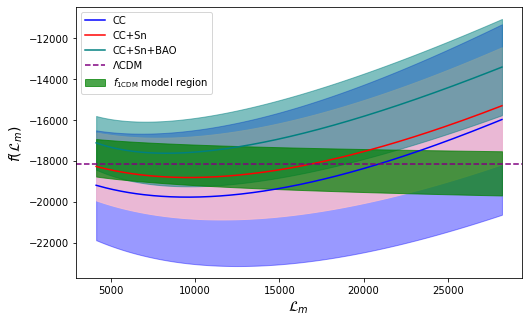}
        \caption{The green shaded region corresponds to the theoretical\\ $f_{1CDM}$ model for $0.018 \leq b_{1} \leq 0.025$.}
        \label{f1(lm)}
    \end{subfigure}%
    \hfill\hfill
    \begin{subfigure}[b]{0.50\textwidth}
        \centering
        \includegraphics[width=88mm]{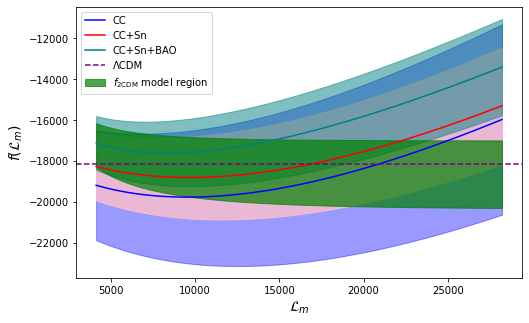}
        \caption{The green shaded region corresponds to the theoretical\\ $f_{2CDM}$ model for $2.3 \leq b_{2} \leq 3.0$.}
        \label{f2(lm)}
    \end{subfigure}%
    \caption{The reconstruction of $f(\mathcal{L}_{m})$ as a function of $\mathcal{L}_{m}$ using GP regression on $H(z)$ and $H'(z)$ for different combinations of observational datasets. The shaded regions represent the $1\sigma$ confidence intervals. The dashed purple line represents the $\Lambda$CDM model.}
    \label{fig:flm reconstruction}
\end{figure}    
In Fig. \ref{fig:flm reconstruction}, we have shown the plot for reconstruction of $f(\mathcal{L}_{m})$ as a function of $\mathcal{L}_{m}$ using the data driven reconstructed values of Hubble and its derivatives in the previous section. We have also plotted the $\Lambda$CDM model depicted by the straight dashed line which lie in the central part of the reconstructed region where $f(\mathcal{L}_{m})=-2\Lambda=-6H_0^2(1-\Omega_{m0})$. The shaded regions represent the $1\sigma$ uncertainties and the solid lines are the mean reconstructed curve. In order to get a appropriate functional form of $f(\mathcal{L}_{m})$ which can describe the reconstructed $f(\mathcal{L}_{m})$, we best-fit the mean curve obtained from combined $CC+SN+BAO$ data with the with a quadratic model $f(\mathcal{L}_{m})=-2\Lambda+\alpha \mathcal{L}_{m}+\zeta \mathcal{L}_{m}^2$ where the value $\alpha \approx -0.08582 \pm 0.00345$ and $\zeta\approx(-1.512584\pm 0.563717)\times {10}^{-6}$. Although we have obtained best fit functional form of the mean reconstructed curve, any model that lies within the $1\sigma$ shaded regions are cosmologically viable so we have considered two $f(\mathcal{L}_{m})$ models \cite{Cai2020}. One is the power-law model $f_{1CDM}$ and the second model as square-root exponential model. In case of $f_{1CDM}=\alpha (\mathcal{L}_{m})^{b_{1}}$ where the value of $\alpha= \frac{1-\Omega_{m0}}{(1+2b_{1})}(6H_0^2)^{(1-b_{1})}$ for $b_{1}$ in the range $0.018 \leq b_{1} \leq 0.025$ is the green shaded region as we can see in Fig. \ref{f1(lm)}. For the square-root exponential model, $f_{2CDM}=\alpha \mathcal{L}_{m0}\left(1-e^{-b_{2}\sqrt{\frac{\mathcal{L}_{m}}{\mathcal{L}_{m0}}}}\right)$ with $\alpha=\frac{1}{[1-(1+b_{2})e^{-b_{2}}]\Omega_{m0}}$. We have plotted for $2.3\leq b_{2} \leq 3.0$ which corresponds to the green shaded region in Fig. \ref{f2(lm)} that lies in the $1\sigma$ reconstructed region.

\section{Dynamical stability analysis} \label{sec5}
In this section, we perform dynamical system analysis for the models reconstructed by the GP in previous section. In cosmology, the dynamics of the Universe is govern by the field equations and to perform dynamical system analysis of any cosmological model we convert the cosmological field equations to autonomous system of equations by defining dimensionless auxiliary variables. In dimensional system analysis, the autonomous system is defined as $x'=f(x)$ \cite{Bahamonde_2018,Mirza_2017}, where $x$ denotes the auxiliary variables, $f(x)$ represents the corresponding autonomous equations and prime denotes the derivative with respect to $N=|\ln a(t)|$. We can evaluate the critical points by evaluating $x'=0$ and the stability of the critical points can be evaluated by calculating the eigenvalues of the Jacobian matrix at the critical point. The signature of the eigenvalues gives us the stability of the critical point, if all the eigenvalues are positive the point is unstable or repeller as the system diverges from that point, it can also be observed in phase space as the trajectories appear to be diverging from the unstable point. If all eigenvalues are negative, it's an attractor, with trajectories converging toward it. Mixed signs indicate a saddle point, where some trajectories approach along the stable point and diverge along the unstable point \cite{Wainwright_Ellis_1997,Coley_1999}. 

For our case, in order to study the stability analysis of the model $f( R,\mathcal{L}_{m})$, we have defined the dimensionless variables of Eq. \eqref{eqn1} and rewritten as,

\begin{equation}\label{dimensionless variables}
x = \frac{\dot{f_R}}{Hf_R}, \quad y = \frac{ f}{6H^2f_R}, \quad z = \frac{R}{6H^2}, \quad u=\frac{f_{\mathcal{L}_{m}}\rho}{3H^2f_R}.
\end{equation}

Using Eq. \eqref{dimensionless variables} in Eq. \eqref{eqn1}, we obtain a constrained equation as $1=z+u-x-y$. Now, by differentiating the variables with dimensionless time variable N we obtain the autonomous system of equations as,
\begin{align}
    x' &= \frac{\ddot{f_R}}{H^2f_R} - x \left( \frac{\dot{H}}{H^2} \right) - x^2 \label{x'eqn}~,\\
    y' &= \frac{\dot{f}}{6H^3f_R} - 2y \left( \frac{\dot{H}}{H^2} \right) - xy \label{y'}~,\\
    z' &= \frac{\dot{R}}{6H^3} - 2z \left( \frac{\dot{H}}{H^2} \right) \label{z'}~,\\
    u' &= -3u \left( 1 + \frac{\rho f_{\mathcal{L}_{m}\mathcal{L}_{m}}}{f_{\mathcal{L}_{m}}} \right) - 2u \left( \frac{\dot{H}}{H^2} \right) - xu \label{u'}~,
\end{align}
For the above equations to be a close system, we must express all the terms in the right-hand side must be expressed in terms of the variables expressed in Eq. \eqref{dimensionless variables}. Using $R=6(\dot{H}+2H^2)$ now we can write,
\begin{align}
    \frac{\dot{H}}{H^2}&=z-2 \label{HdotH^2}~,\\
    \frac{\dot{R}}{6H^3}&=\frac{xz}{b} \quad where \quad b=\frac{d\ln f_R}{d\ln R}=\frac{f_{RR}R}{f_R} \label{Rdot6H^3}~,\\
    \frac{\dot{f}}{6H^3f_R}&=\frac{xz}{b}-3u \label{fdot6H^3fR}~.\\
\end{align}
The above equations govern the cosmological evolution of the Universe for a generalized $f(R,\mathcal{L}_m)$ gravity theory specified by $\Gamma=\frac{\ddot{f_R}}{H^2f_R}$ and $\lambda=\frac{\rho f_{\mathcal{L}_m \mathcal{L}_m}}{f_{\mathcal{L}_m}}$. For the power-law model $f(R,\mathcal{L}_m)=\frac{R}{2}+\alpha \mathcal{L}_m^{b_1}$ the variable $x=0$ and $y=z+\frac{u}{b_1}$ so the autonomous close independent system of equations becomes
\begin{eqnarray}
    \label{model1z'}
    z'&=&-2z(z-2)~,\\
    u'&=&-3u(1+\lambda)-2u(z-2)~,~~\text{where}~~\lambda =(b_1-1)~.\nonumber\\
        \label{model1u'}
\end{eqnarray}

\begin{table}[ht]
\centering
\resizebox{\columnwidth}{!}{%
\begin{tabular}{|c|c|c|c|c|c|c|}
\hline
critical point & \(z\) & \(u\) & Eigen values &\(q(z)\) & \(w_{\mathrm{eff}}(z)\) & Stability \\ 
\hline
A & 0 & 0 & (4,3.925) &1  & \(\tfrac{1}{3}\)  & unstable \\ 
\hline
B & 2 & 0 & (-4,-0.075)& \(-1\) & \(-1\) & stable \\ 
\hline
\end{tabular}
}
\caption{critical points and corresponding deceleration and effective-EoS parameters for power-law model.}
\label{tab:crit_points Model-I}
\end{table}

\begin{figure}[ht]
\centering
\includegraphics[width=0.5\textwidth]{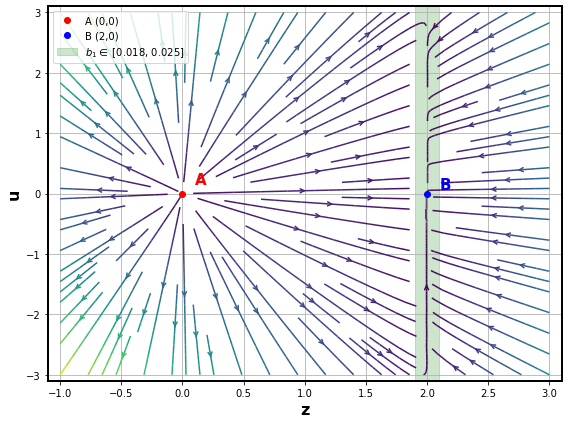}
\caption{Phase space portrait for the power-law model in the $z-u$ plane, with the viable range of model parameter in the stable green shaded region.}
\label{fig:phase_space Model-I}
\end{figure}
From Table \ref{tab:crit_points Model-I} and Fig. \ref{fig:phase_space Model-I} we can understand the stability and cosmological behavior of the critical points obtained as:\\
critical point A$(0, 0)$, corresponds to a flat, curvature-free Universe with vanishing effective matter density. The deceleration parameter at this point is $q = 1$ and the effective equation of state is $w_{\text{eff}} = \frac{1}{3}$, which resembles a radiation-dominated phase. However, since the matter contribution $u$ is zero, this state does not represent a realistic cosmological era. Instead, it may be interpreted as a vacuum-like or early pre-matter phase. The positive eigenvalues of the linearized system $(4,~3.925)$ indicate that the point is an unstable node and the Universe evolves away from this configuration. Thus, critical point A plays the role of a past-time repeller, marking an initial state from which the cosmological evolution proceeds.\\
critical point B$(2, 0)$, represents a late-time cosmological state characterized by constant curvature and a vanishing effective matter density. At this point, the deceleration parameter is $q = -1$ and the effective equation of state is $w_{\text{eff}} = -1$, consistent with an accelerating de Sitter-like expansion. The eigenvalues $(-4,~-0.075)$ confirm that this point is a stable attractor, suggesting that the Universe naturally evolves toward this configuration at late times. \\

Now for the square-root-exponential model $f(R,\mathcal{L}_m)=\frac{R}{2}+\alpha \mathcal{L}_{m0}\left(1-e^{-b_{2}\sqrt{\frac{\mathcal{L}_{m}}{\mathcal{L}_{m0}}}}\right)$, the dynamical variables and autonomous equations defined above in Eq.~\eqref{dimensionless variables} - Eq.~\eqref{fdot6H^3fR}, remain the same and the term $\lambda=\frac{\rho f_{\mathcal{L}_m\mathcal{L}_{m}}}{f_{\mathcal{L}_m}}=-\left( \frac{\beta}{2} \sqrt{\frac{\rho}{\rho 0}}+\frac{1}{2}\right)$, so we define a new variable $s=\sqrt{\frac{\rho}{\rho_0}}$ and can also be expressed as $y=u+s\big(\alpha s-\frac{y}{\beta}\big)$ so the final autonomous system of equations becomes
\begin{eqnarray}
    \label{model2z'}
    z'&=&-2z(z-2)~,\\
    \label{models'}
    s'&=&-\frac{3}{2}s~,\\
    u'&=&-3u(1+\lambda)-2u(z-2),~\text{where}~ \lambda =-(\frac{b_2}{2}s+\frac{1}{2})~.\nonumber\\
        \label{model2u'}
\end{eqnarray}

\begin{table}[ht]
\centering
\resizebox{\columnwidth}{!}{%
\begin{tabular}{|c|c|c|c|c|c|c|c|}
\hline
critical point & \(z\) & \(u\) & \(s\) & Eigen values & \(q(z)\) & \(w_{\mathrm{eff}}(z)\) & Stability \\ 
\hline
P$_{1}$ & 0 & 0 & 0 & (4.0,-1.5,-5.5) & $1$  & $\frac{1}{3}$  & saddle  \\[3pt] 
\hline
P$_{2}$ & 2 & 0 & 0 & (-4,-1.5,-1.5)& \(-1\) & \(-1\) & stable \\ 
\hline
\end{tabular}
}
\caption{critical points and corresponding deceleration and effective-EoS parameters.}
\label{tab:crit_points Model-II}
\end{table}

\begin{figure}[ht]
\centering
\includegraphics[width=0.5\textwidth]{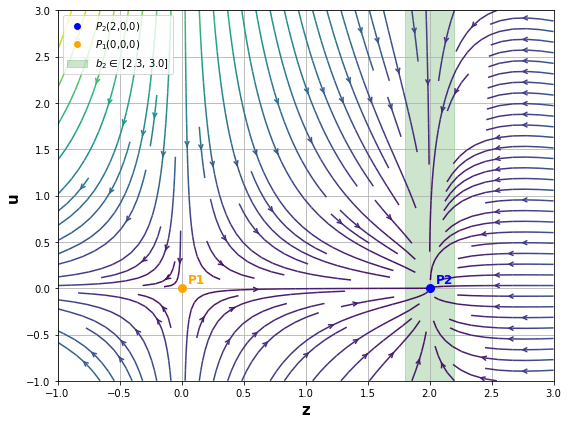}
\caption{Phase space portrait for the square-root-exponential model in the $z-u$ plane at fix $s=0$, with the viable range of model parameter in the stable green shaded region.}
\label{fig:phase_space Model-II}
\end{figure}

critical point P$_{1} (0,~0,~0)$ corresponds to an early-time configuration where both the matter content and normalized energy density vanish. The deceleration parameter at this point is $q = 1$, consistent with a decelerating Universe similar to a radiation-dominated phase. However, since the actual matter energy density is zero, this configuration represents a transient, non-physical phase. The eigenvalues of the Jacobian matrix are $(4.0,~-1.5,~-5.5)$, indicating a saddle point. The mixed sign of eigenvalues implies that the system may evolve toward this point along certain directions but will eventually move away from it. \\
critical point P$_{2} (2,~0,~0)$, this point represents a stable late-time attractor. The curvature variable $z = 2$ yields a deceleration parameter $q = -1$ and the effective equation of state becomes $w_{\text{eff}} = -1$, matching the expected behavior of a dark energy–dominated, de Sitter Universe. The eigenvalues at this point are all negative, specifically $(-4.0,~-1.5,~-1.5)$, which confirms that this is a stable node. The phase space trajectories converge toward this point, indicating that the Universe naturally evolves toward this state at late times.

\section{Conclusion}\label{sec6}
In this study, we have independently reconstructed the functional form of $f(\mathcal{L}_m)$ using GP regression within the framework of $f(R, \mathcal{L}_m)$ gravity. We consider a separable form of the action given by $f(R, \mathcal{L}_m) = g(R) + f(\mathcal{L}_m)$, where $g(R) = \frac{R}{2}$ ensures that the geometric part reduces to GR in the appropriate limit. To perform a model independent reconstruction, we utilized observational datasets from CC, SN and BAO and applied GP techniques to extract both the Hubble parameter and its derivative. Using these reconstructed quantities and assuming the matter Lagrangian $\mathcal{L}_m = \rho$, we reconstructed the functional form $f(\mathcal{L}_m)$. A comparative analysis with the standard $\Lambda$CDM model reveals that the $\Lambda$CDM curve lies within the 1$\sigma$ confidence region of the reconstructed mean curve across all datasets. However, the deviation of the mean reconstructed curves from the $\Lambda$CDM trajectory indicates a quadratic behavior in $f(\mathcal{L}_m)$. Motivated by this, we propose the following functional form $f(\mathcal{L}_m) = -2\Lambda + \alpha \mathcal{L}_m + \zeta \mathcal{L}_m^2$ and constrain the parameters to be $\alpha \approx -0.08582 \pm 0.00345$ and $\zeta \approx (-1.512584 \pm 0.563717) \times 10^{-6}$. In addition, we explore two alternative models that also lie within the 1$\sigma$ region of the reconstructed curve as (i) power-law model: $f_{1\text{CDM}} = \alpha \mathcal{L}_m^{b_1}$, where the exponent lies within the range $0.018 \leq b_1 \leq 0.025$ and (ii) square-root-exponential model: $f_{2\text{CDM}} = \alpha \mathcal{L}_{m0} \left(1 - e^{-\beta \sqrt{\mathcal{L}_m / \mathcal{L}_{m0}}} \right)$, with $2.3 \leq \beta \leq 3.0$. These results support the viability of non-minimal matter-curvature coupling and open up new avenues for modified gravity theories beyond the standard cosmological model.

The dynamical system analysis has been performed for both the cosmological models obtained with the form of $f(\mathcal{L}_{m})$. For the the power-low model we get two critical point A and B, critical point A $(0,0)$ shows early unstable behavior and critical point B $(2,0)$ shows late time stable dark energy dominating era. The point A is an unstable node with all eigenvalues for Jacobian matrix positive real part, and the point B has all eigenvalues for Jacobian matrix are negative real part and zero gives us the stable node behavior. For more ideas about the critical point, the $2-D$ phase portrait has been shown in Fig. \ref{fig:phase_space Model-I}. Similarly for square-root-exponential model, the critical points are P$_{1}$ $(0, 0, 0)$ and P$_{2}$ $(2,0,0)$. Here also, $P_{1}$ is unstable point, whereas $P_{2}$ shows the stable node behavior. Both the models shows early radiation era and late time DE era.\\

In summary, our results demonstrate that it is possible to reconstruct the functional form of $f(\mathcal{L}_m)$ in a completely data-driven and model-independent manner using GP and current observational Hubble data. The reconstructed function deviates from the $\Lambda$CDM behavior, motivating the consideration of alternative models such as the quadratic, power law and square-root exponential forms. These models not only remain consistent with the 1$\sigma$ reconstruction but also yield cosmologically viable dynamics, as confirmed through dynamical system analysis. The presence of stable attractor solutions at late times for both models supports their potential relevance in describing the cosmic acceleration without explicitly invoking a cosmological constant. This approach provides a powerful framework for testing the viability of modified gravity models directly from observations and opens the door for future explorations with more precise datasets and extended theoretical forms of $f(R, \mathcal{L}_m)$ gravity.

\section*{Acknowledgment}  BM acknowledges the support of Council of Scientific and Industrial Research (CSIR) for the project grant (No. 03/1493/23/EMR II)

\section*{References}
\bibliographystyle{JCAP.bst}
\bibliography{references}

\providecommand{\href}[2]{#2}\begingroup\raggedright\begin{thebibliography}{10}

\bibitem{Riess_1998}
A.G.~Riess, A.V.~Filippenko, P.~Challis, A.~Clocchiatti et~al.,
  \emph{Observational evidence from supernovae for an accelerating universe and
  a cosmological constant}, \href{https://doi.org/10.1086/300499}{\emph{The
  Astronomical Journal} {\bfseries 116} (1998) 1009}.

\bibitem{Perlmutter_1999}
S.~Perlmutter, G.~Aldering, G.~Goldhaber, R.A.~Knop et~al., \emph{Measurements
  of $\omega$ and $\lambda$ from 42 high-redshift supernovae},
  \href{https://doi.org/10.1086/307221}{\emph{The Astrophysical Journal}
  {\bfseries 517} (1999) 565}.

\bibitem{Spergel_2007}
D.N.~Spergel, R.~Bean, O.~Doré, M.R.~Nolta et~al., \emph{Three-year wilkinson
  microwave anisotropy probe (wmap) observations: Implications for cosmology},
  \href{https://doi.org/10.1086/513700}{\emph{Astrophysical Journal Supplement
  Series} {\bfseries 170} (2007) 377}.

\bibitem{Komatsu_2011}
E.~Komatsu, K.M.~Smith, J.~Dunkley, C.L.~Bennett et~al., \emph{Seven-year
  wilkinson microwave anisotropy probe (wmap) observations: Cosmological
  interpretation},
  \href{https://doi.org/10.1088/0067-0049/192/2/18}{\emph{Astrophysical Journal
  Supplement Series} {\bfseries 192} (2011) 18}.

\bibitem{Eisenstein_2005}
D.J.~Eisenstein, I.~Zehavi, D.W.~Hogg, R.~Scoccimarro et~al., \emph{Detection
  of the baryon acoustic peak in the large-scale correlation function of sdss
  luminous red galaxies}, \href{https://doi.org/10.1086/466512}{\emph{The
  Astrophysical Journal} {\bfseries 633} (2005) 560}.

\bibitem{Percival_2010}
W.J.~Percival, B.A.~Reid, D.J.~Eisenstein, N.A.~Bahcall et~al., \emph{Baryon
  acoustic oscillations in the sloan digital sky survey data release 7 galaxy
  sample},
  \href{https://doi.org/10.1111/j.1365-2966.2009.15812.x}{\emph{Monthly Notices
  of the Royal Astronomical Society} {\bfseries 401} (2010) 2148}.

\bibitem{Weinberg_2013}
D.H.~Weinberg, M.J.~Mortonson, D.J.~Eisenstein, C.~Hirata et~al.,
  \emph{Observational probes of cosmic acceleration},
  \href{https://doi.org/10.1016/j.physrep.2013.05.001}{\emph{Physics Reports}
  {\bfseries 530} (2013) 87}.

\bibitem{Weinberg1989}
S.~Weinberg, \emph{The cosmological constant problem},
  \href{https://doi.org/10.1103/RevModPhys.61.1}{\emph{Reviews of Modern
  Physics} {\bfseries 61} (1989) 1}.

\bibitem{Wilson2006}
K.M.~Wilson, G.~Chen and B.~Ratra, \emph{Supernova ia and galaxy cluster gas
  mass fraction constraints on dark energy},
  \href{https://doi.org/10.1142/S0217732306023180}{\emph{Modern Physics Letters
  A} {\bfseries 21} (2006) 2197}.

\bibitem{Davis2007}
T.M.~Davis, E.~M\"ortsell, J.~Sollerman, A.C.~Becker et~al., \emph{Scrutinizing
  exotic cosmological models using essence supernova data combined with other
  cosmological probes}, \href{https://doi.org/10.1086/519988}{\emph{The
  Astrophysical Journal} {\bfseries 666} (2007) 716}.

\bibitem{Steinhardt1999}
P.J.~Steinhardt, L.~Wang and I.~Zlatev, \emph{Cosmological tracking solutions},
  \href{https://doi.org/10.1103/PhysRevD.59.123504}{\emph{Physical Review D}
  {\bfseries 59} (1999) 123504}.

\bibitem{Copeland2006}
E.J.~Copeland, M.~Sami and S.~Tsujikawa, \emph{Dynamics of dark energy},
  \href{https://doi.org/10.1142/S021827180600942X}{\emph{International Journal
  of Modern Physics D} {\bfseries 15} (2006) 1753}.

\bibitem{Padmanabhan2003}
T.~Padmanabhan, \emph{Cosmological constant—the weight of the vacuum},
  \href{https://doi.org/10.1016/S0370-1573(03)00120-0}{\emph{Physics Reports}
  {\bfseries 380} (2003) 235}.

\bibitem{CAPOZZIELLO2011167}
S.~Capozziello and M.~{De Laurentis}, \emph{Extended theories of gravity},
  \href{https://doi.org/https://doi.org/10.1016/j.physrep.2011.09.003}{\emph{Physics
  Reports} {\bfseries 509} (2011) 167}.

\bibitem{NOJIRI201159}
S.~Nojiri and S.D.~Odintsov, \emph{Unified cosmic history in modified gravity:
  From f(r) theory to lorentz non-invariant models},
  \href{https://doi.org/https://doi.org/10.1016/j.physrep.2011.04.001}{\emph{Physics
  Reports} {\bfseries 505} (2011) 59}.

\bibitem{Sotiriou2010}
T.P.~Sotiriou and V.~Faraoni, \emph{${f(R)}$ theories of gravity},
  \href{https://doi.org/10.1103/RevModPhys.82.451}{\emph{Reviews of Modern
  Physics} {\bfseries 82} (2010) 451}.

\bibitem{Harko2010}
T.~Harko, F.S.N.~Lobo, S.~Nojiri and S.D.~Odintsov, \emph{${f(R, L_{m})}$
  gravity}, \href{https://doi.org/10.1103/PhysRevD.81.084050}{\emph{Physical
  Review D} {\bfseries 81} (2010) 084050}.

\bibitem{Chevallier2001}
M.~Chevallier and D.~Polarski, \emph{Accelerating universes with scaling dark
  matter}, \href{https://doi.org/10.1142/S0218271801000822}{\emph{International
  Journal of Modern Physics D} {\bfseries 10} (2001) 213}.

\bibitem{Linden2008}
S.~Linden and J.-M.~Virey, \emph{Test of the chevallier-polarski-linder
  parametrization for rapid dark energy equation of state transitions},
  \href{https://doi.org/10.1103/PhysRevD.78.023526}{\emph{Physical Review D}
  {\bfseries 78} (2008) 023526}.

\bibitem{Visser2004}
M.~Visser, \emph{Jerk, snap and the cosmological equation of state},
  \href{https://doi.org/10.1088/0264-9381/21/11/006}{\emph{Classical and
  Quantum Gravity} {\bfseries 21} (2004) 2603}.

\bibitem{Rasmussen2005}
C.E.~Rasmussen and C.K.I.~Williams, \emph{Gaussian Processes for Machine
  Learning}, Adaptive Computation and Machine Learning, The MIT Press (2005),
  \href{https://doi.org/10.7551/mitpress/3206.001.0001}{10.7551/mitpress/3206.001.0001}.

\bibitem{Seikel2013}
M.~Seikel and C.~Clarkson, \emph{Optimising gaussian processes for
  reconstructing dark energy dynamics from supernovae},  2013.

\bibitem{Seikel_2012}
M.~Seikel, C.~Clarkson and M.~Smith, \emph{{Reconstruction of dark energy and
  expansion dynamics using Gaussian processes}},
  \href{https://doi.org/10.1088/1475-7516/2012/06/036}{\emph{J. Cosmol.
  Astropart. Phys.} {\bfseries 2012} (2012) 036}.

\bibitem{Shafieloo2014}
A.~Shafieloo, \emph{Falsifying cosmological constant},
  \href{https://doi.org/10.1016/j.nuclphysbps.2013.10.081}{\emph{Nuclear
  Physics B - Proceedings Supplements} {\bfseries 246--247} (2014) 171}.

\bibitem{Kim2013}
A.G.~Kim, R.C.~Thomas, G.~Aldering, P.~Antilogus et~al., \emph{Standardizing
  type ia supernova absolute magnitudes using gaussian process data
  regression}, \href{https://doi.org/10.1088/0004-637X/766/2/84}{\emph{The
  Astrophysical Journal} {\bfseries 766} (2013) 84}.

\bibitem{Nair2014}
R.~Nair, S.~Jhingan and D.~Jain, \emph{Reconstruction of dark energy
  interaction from observational data},
  \href{https://doi.org/10.1088/1475-7516/2014/01/005}{\emph{Journal of
  Cosmology and Astroparticle Physics} {\bfseries 2014} (2014) 005}.

\bibitem{Yang2015}
T.~Yang, Z.-K.~Guo and R.-G.~Cai, \emph{Reconstructing the interaction between
  dark energy and dark matter using gaussian processes},
  \href{https://doi.org/10.1103/PhysRevD.91.123533}{\emph{Physical Review D}
  {\bfseries 91} (2015) 123533}.

\bibitem{GomezValent2018}
A.~Gómez-Valent and L.~Amendola, \emph{H$_{0}$ from cosmic chronometers and
  type ia supernovae, with gaussian processes and the novel weighted polynomial
  regression method},
  \href{https://doi.org/10.1088/1475-7516/2018/04/051}{\emph{Journal of
  Cosmology and Astroparticle Physics} {\bfseries 2018} (2018) 051}.

\bibitem{Cai2020}
Y.-F.~Cai, M.~Khurshudyan and E.N.~Saridakis, \emph{Model-independent
  reconstruction of f(t) gravity from gaussian processes},
  \href{https://doi.org/10.3847/1538-4357/ab5a7f}{\emph{The Astrophysical
  Journal} {\bfseries 888} (2020) 62}.

\bibitem{Moresco_2016}
M.~Moresco, L.~Pozzetti, A.~Cimatti, R.~Jimenez et~al., \emph{A 6\% measurement
  of the hubble parameter at \(z \sim 0.45\): direct evidence of the epoch of
  cosmic re-acceleration},
  \href{https://doi.org/10.1088/1475-7516/2016/05/014}{\emph{J. Cosmol.
  Astropart. Phys.} {\bfseries 2016} (2016) 014}.

\bibitem{Riess_2018}
A.G.~Riess, S.A.~Rodney, D.M.~Scolnic and D.L.~Shafer, \emph{Type ia supernova
  distances at redshift $>$1.5 from the hubble space telescope multi-cycle
  treasury programs: The early expansion rate},
  \href{https://doi.org/10.3847/1538-4357/aaa5a9}{\emph{Astrophys. J.}
  {\bfseries 853} (2018) 126}.

\bibitem{GomezValent_2018}
A.~G\'omez-Valent and L.~Amendola, \emph{{$H_0$ from cosmic chronometers and
  Type Ia supernovae, with Gaussian Processes and the novel Weighted Polynomial
  Regression method}},
  \href{https://doi.org/10.1088/1475-7516/2018/04/051}{\emph{J. Cosmol.
  Astropart. Phys.} {\bfseries 2018} (2018) 051}.

\bibitem{Zhang_2016}
M.-J.~Zhang and J.-Q.~Xia, \emph{{Test of the cosmic evolution using Gaussian
  processes}}, \href{https://doi.org/10.1088/1475-7516/2016/12/005}{\emph{J.
  Cosmol. Astropart. Phys.} {\bfseries 2016} (2016) 005}.

\bibitem{Harko_2015}
T.~Harko, F.S.N.~Lobo, J.P.~Mimoso and D.~Pav{\'o}n, \emph{Gravitational
  induced particle production through a nonminimal curvature--matter coupling},
  \href{https://doi.org/10.1140/epjc/s10052-015-3517-4}{\emph{Eur. Phys. J. C}
  {\bfseries 75} (2015) 1}.

\bibitem{Planck_2020}
N.~Aghanim, Y.~Akrami, M.~Ashdown and J.~Aumont, \emph{Planck 2018 results: Vi.
  cosmological parameters},
  \href{https://doi.org/10.1051/0004-6361/201833910}{\emph{Astron. Astrophys.}
  {\bfseries 641} (2020) A6}.

\bibitem{Bahamonde_2018}
S.~Bahamonde, C.G.~Böhmer, S.~Carloni, E.J.~Copeland et~al., \emph{Dynamical
  systems applied to cosmology: Dark energy and modified gravity},
  \href{https://doi.org/10.1016/j.physrep.2018.09.001}{\emph{Physics Reports}
  {\bfseries 775-777} (2018) 1}.

\bibitem{Mirza_2017}
B.~Mirza and F.~Oboudiat, \emph{Constraining ${f(T)}$ gravity by dynamical
  system analysis},
  \href{https://doi.org/10.1088/1475-7516/2017/11/011}{\emph{Journal of
  Cosmology and Astroparticle Physics} {\bfseries 2017} (2017) 011}.

\bibitem{Wainwright_Ellis_1997}
J.~Wainwright and G.F.R.~Ellis, \emph{Dynamical Systems in Cosmology},
  Cambridge University Press (1997).

\bibitem{Coley_1999}
A.A.~Coley, \emph{Dynamical systems in cosmology}, {\emph{arXiv preprint
  gr-qc/9910074} (1999) }
  [\href{https://arxiv.org/abs/gr-qc/9910074}{{\ttfamily gr-qc/9910074}}].

\end{thebibliography}\endgroup

\end{document}